\documentclass{PoS}

\def\beq{\begin{equation}}
\def\eeq{\end{equation}}
\def\bea{\begin{eqnarray}}
\def\eea{\end{eqnarray}}

\def\ep{\epsilon}

\def\nn{\nonumber}
\def\Eq#1{Eq.~(\ref{#1})}
\def\ln#1{\mathrm{log}\left(#1\right)}

\def\li#1{\mathrm{Li_2}\left(#1\right)}

\def\td#1{\tilde{\delta}\left(#1\right)}

\def\cgt{\widetilde{c}_{\Gamma}}
\newcommand{\la}{\langle}
\newcommand{\ra}{\rangle}

\def\qb{\mathbf{q}}

\def\gs{g_{\rm S}}
\def\v{{\rm V}}
\def\r{{\rm R}}

\newcommand\as{\alpha_{\mathrm{S}}}

\title{
\vspace*{-2.5cm}
\begin{minipage}{\textwidth}
{\normalfont\small IFIC/15-71
\hspace{\fill} January 2016
}\\
\end{minipage}\\[60pt]
  Loop-tree duality and quantum field theory in four dimensions}

\ShortTitle{LTD and quantum field theory in 4D}

\author{\speaker{Germ\'an F. R. Sborlini}$^{a,b}$\\\\
        $^a$Instituto de F\'{\i}sica Corpuscular, Universitat de Val\`{e}ncia -- 
Consejo Superior de Investigaciones Cient\'{\i}ficas, 
Parc Cient\'{\i}fic, E-46980 Paterna, Valencia, Spain. \\\
				$^b$Departamento de F\'\i sica and IFIBA, FCEyN, Universidad de Buenos Aires, 
(1428) Pabell\'on 1 Ciudad Universitaria, Capital Federal, Argentina. \\\\
        E-mail: \email{gfsborlini@df.uba.ar}}

\abstract{Loop-tree duality allows to express virtual contributions in terms of phase-space integrals, thus leading to a direct comparison with real radiation terms. In this talk, we review the basis of the method and describe its application to regularize Feynman integrals. Performing an integrand-level combination of real and virtual terms, we obtain finite contributions that can be computed in four-dimensions. Moreover, this method provides a natural physical interpretation of infrared singularities, their origin and the way that they cancel in the complete computation.}

\FullConference{12th International Symposium on Radiative Corrections (Radcor 2015) and LoopFest XIV (Radiative Corrections for the LHC and Future Colliders)\\15-19 June, 2015\\UCLA Department of Physics $\&$ Astronomy, Los Angeles, USA}

\begin{document}

\section{Introduction}
\label{sec:introduction}
The knowledge of higher-order corrections is crucial for the discovery and characterization of potential new-physics signals. From the theoretical point of view, this is a very challenging task due to the increasing computational complexity of multi-loop multi-leg processes. In fact, the presence of singularities (or \emph{ill-defined} expressions) forces the application of regularization methods to obtain finite results, such as dimensional regularization (DREG) \cite{Bollini:1972ui, 'tHooft:1972fi, Cicuta:1972jf, Ashmore:1972uj}. There are different kind of singularities;
\begin{itemize}
\item ultraviolet (UV), associated with the high-energy behavior of the theory;
\item infrared (IR), related with the presence of degenerate configurations in the low-energy limit; 
\item and threshold singularities, which may appear when virtual particles are produced on-shell; 
\end{itemize}
as well as other spurious or non-physical divergences. In the context of DREG, threshold singularities do not introduce $\epsilon$-poles because they are integrable singularities; however, it is crucial to introduce a proper \emph{prescription} to deal with them. On the other hand, renormalization successfully cures UV divergences by applying a well-known and systematic procedure. For the IR case, it is also possible to achieve a finite result when IR-safe observables are considered. Only under that assumption, Kinoshita-Lee-Nauenberg (KLN) \cite{Kinoshita:1962ur} theorem guarantees the cancellation of IR singularities among all the possible degenerate configurations associated with the same final state observable. In particular, this involves taking into account real and virtual corrections, which differ in the number of external particles.

In this article, we focus in the treatment of IR singularities through a new method, which is based in the loop-tree duality (LTD) theorem \cite{Catani:2008xa, Hernandez-Pinto:2015ysa,Sborlini:2015uia}. In the standard approach, which is based in the subtraction method \cite{Catani:1996vz,Catani:1996jh,Frixione:1995ms,GehrmannDeRidder:2005cm,Catani:2007vq,DelDuca:2015zqa,Czakon:2010td,Boughezal:2015dva,Gaunt:2015pea}, the IR-singular behavior of scattering amplitudes is exploited to write local counter-terms which could be integrated analytically. The integrated counter-terms are combined with the virtual contributions, whilst their unintegrated form is used to perform a local regularization of the real terms. Contrary to this, the main idea behind LTD is to rewrite virtual amplitudes as real radiation terms, and perform an integrand level combination with the real contribution. In this way, dual amplitudes originated from the virtual part act directly as counter-terms for the real-emission amplitudes.

The outline of this article is the following. In Section \ref{sec:LTD} we review the basic ideas behind LTD, introducing suitable notation. Then, in Section \ref{sec:exampleFeynman}, we study the divergent structure of a triangle Feynman integral. In particular, we prove that IR poles are originated in a compact region of the integration space. After that, we explain the implementation of a fully inclusive cross-section computation at next-to-leading order (NLO) in Section \ref{sec:examplephysical}. We focus the discussion in the real-virtual momenta mapping, which plays a crucial role to achieve the IR cancellation. Some details about the numerical implementation are given, emphasizing the possibility of obtaining four-dimensional representations at integrand level. Finally, in Section \ref{sec:conclusions} we present the conclusions and discuss about possible extensions of these ideas.


\section{Basics of LTD}
\label{sec:LTD}
The real and virtual contributions share the same kind of IR singularities, in spite of involving a different number of final-state particles. In LTD loop integrals are related with phase-space integrals of tree-level objects called \emph{dual integrals} \cite{Catani:2008xa}. In any relativistict, local and unitary theory, this property is directly translated to scattering amplitudes, and virtual amplitudes become expresible in terms of dual amplitudes. In this article, we analyze one-loop corrections, but these ideas are naturally extended to higher-loops \cite{Catani:2008xa,Bierenbaum:2010cy,Bierenbaum:2012th,Buchta:2014dfa,Buchta:2015xda}. Let's consider an $N$-leg scalar one-loop integral, whose dual representation is given by the sum of $N$ dual integrals. Each dual integral is associated with a possible one-cut, so we have
\bea
L^{(1)}(p_1, \dots, p_N) 
&=& - \sum_{i\in \alpha_1} \prod_{j \in \alpha_1, \, j\neq i} \, \int_{\ell}  \, \td{q_i} \,G_D(q_i;q_j)~,
\label{eq:LTDoneloop}
\eea 
where the cut condition is obtained through
\beq
\td{q_i} \equiv 2 \pi \, \imath \, \theta(q_{i,0}) \, \delta(q_i^2-m_i^2) \, ,
\label{eq:LTDoneloopDELTA}
\eeq
which forces $q_i$ to be on-shell, and
\beq
G_D(q_i;q_j) = \frac{1}{q_j^2 -m_j^2 - \imath 0 \, \eta \cdot k_{ji}} \ \ \ \ \ \ i,j \in \alpha_1 = \{ 1,2,\ldots N\} \, \, ,
\label{eq:LTDoneloopPROPAGATORDUAL}
\eeq
are dual propagators, with $k_{ji}=q_j-q_i$. The four-momenta of the external legs are $p_{i}$, which are taken as outgoing, and we use $\ell$ as the loop momentum. Also, we denote the internal line momenta as $q_{i,\mu} = (q_{i,0},\mathbf{q}_i)$, where $q_{i,0}$ is the energy component whilst $\qb_{i}$ refers to the spacial part. If $k_{i} = p_{1} + \ldots + p_{i}$ represents a sum of external momenta (which fulfills $k_{N}=0$ due to momentum conservation), then we have $q_{i}=\ell +k_i$ and $k_{ji}=k_j-k_i$.

The distinctive feature of LTD is the introduction of the modified $\imath 0$ prescription. The idea of using the residue theorem and cut diagrams was already applied in the Feynman's tree theorem (FTT) \cite{Feynman:1963ax,Feynman:1972mt}, which establishes that loop amplitudes are reconstructed as the sum over all possible multiple-cuts. In fact, these $m$-cuts are simply obtained by replacing propagators with the associated Dirac's delta given in \Eq{eq:LTDoneloopDELTA}; no other modification is required to recover the exact loop integral. On the other hand, LTD only makes use of single cuts by modifying the Feynman's prescription and introducing the arbitrary future-like vector $\eta$. Notice that the prescription depends on the sign of the product $\eta \cdot k_{ij}$, i.e. different cuts might have a different prescription. This point is crucial to exactly recover the discontinuity structure of virtual amplitudes. In other terms, since FTT and LTD are equivalent, the multiple-cut information is codified into the dual propagators through their modified $\imath 0$ prescription.

Finally, let's make a comment about the relation among loop and phase-space (PS) measures. In the context of DREG, virtual contributions involve the integral over $\ell$ without any other constraint. On the other hand, real radiation terms are obtained after the integration over the extra particle's PS. Since real particles are involved, they are subjected to physical requirements, i.e. they must fulfill momentum conservation and the corresponding on-shell relation. The last condition is implemented through the introduction of the Dirac's delta given in \Eq{eq:LTDoneloopDELTA}. In consequence, we can start from the virtual contribution and use LTD at one-loop; the dual integration measure becomes
\beq
\int_{\ell} \, \td{q_i} = - \imath \mu^{4-d} \int \frac{d^d \ell}{(2\pi)^{d}} \td{q_i} ~, 
\eeq
which resembles a real PS measure in $d$-dimensions. So, LTD converts the usual $d$-dimensional loop measure into a $(d-1)$-dimensional integration. The associated integration domain becomes the forward on-shell hyperboloid associated with the solution of the equation $G_F(q_i)^{-1}=(q_i^2-m_i^2+\imath 0)=0$. In this way, dual contributions are expressed in the same form as the real part, allowing a direct combination at integrand level. We will explain the consequences of this fact in the following sections.


\section{Formal example: disentangling IR singularities in loops}
\label{sec:exampleFeynman}
LTD offers the attractive possibility of studying the divergent structure of virtual contributions (loop integrals, in particular) and identifying the regions responsible of the appearance of these singularities. Let's analyze the simplest IR-divergent Feynman integral, i.e. a scalar triangle in the time-like (TL) region. The process is represented by the kinematical configuration $p_3 \to p_1 + p_2$, with $p_1^2=0=p_2^2$ and $p_3^2=s_{12}>0$. Using the notation introduced in Section \ref{sec:LTD}, we have $q_1 = \ell + p_1$, $q_2 = \ell + p_{12}$ and $q_3 = \ell$. Then, we parametrize $q_i$ in the center-of-mass frame, choosing $\mathbf{p}_1$ ($\mathbf{p}_2$) along the positive (negative) $z$-axis. Using the dimensionless variables $(\xi_{i,0},v_i)$, the kinematical invariants of this system are given by
\beq
2 q_i\cdot p_1 / s_{12} = \xi_{i,0}\, v_i \, , \ \ \ \ \ 2 q_i\cdot p_2 / s_{12} = \xi_{i,0}\, (1-v_i) \, ,
\eeq
and we define the $d$-dimensional integration measures as 
\beq
d[\xi_{i,0}] = c_{\Gamma} \,
\left(\frac{s_{12}}{\mu^2}\right)^{-\ep} \, \xi_{i,0}^{-2\ep} \, d\xi_{i,0}~, \ \ \ \ d[v_i] = (v_i(1-v_i))^{-\ep} \, dv_i~,
\eeq
with $c_{\Gamma}$ the usual loop volume factor in $d$ dimensions. Here, $\xi_{i,0}$ is related to the energy component of $q_i$, whilst $v_i$ is an angular variable. Since there are three internal lines, the application of LTD leads to 
\bea
L^{(1)}(p_1,p_2,-p_3) &=& \frac{c_{\Gamma}}{s_{12}\,\epsilon^2} \left(\frac{-s_{12}-\imath 0}{\mu^2}\right)^{-\epsilon} \ = \sum_{i=1}^3 I_{i}~\, ,
\label{triangulo}
\eea
with the dual integrals
\bea
I_1 &=& \frac{1}{s_{12}} \, \int_{0}^{\infty} d[\xi_{1,0}] \, \int_{0}^{1} d[v_1] \,  
\xi_{1,0}^{-1} \, (v_1 (1-v_1))^{-1}~, 
\label{dualintegrals1}
\\ I_2 &=& \frac{1}{s_{12}} \, \int_{0}^{\infty} d[\xi_{2,0}] \, \int_{0}^{1} d[v_2] \,  
\frac{(1-v_2)^{-1}}{1 - \xi_{2,0} + \imath 0}~, 
\label{dualintegrals2}
\\ I_3 &=& - \frac{1}{s_{12}} \, \int_{0}^{\infty} d[\xi_{3,0}] \, \int_{0}^{1} [v_3] \,  
\frac{v_3^{-1}}{1 + \xi_{3,0}}~.
\label{dualintegrals3}
\eea
It is worth appreciating that the dual prescription becomes relevant for $I_2$, because the denominator vanishes inside the integration domain. This is related with the presence of a threshold singularity. Explicitly, two internal lines become simultaneously on-shell when $\xi_{2,0}=1$ and the diagram can be split into two physical tree-level terms. These contributions resemble those obtained after the application of Cutkowsky's rules. In fact, we can remove the imaginary part of $I_2$ by adding the Cutkowsky contribution, and this removes the purely imaginary $\epsilon$-poles allowing to obtain a four-dimensional representation of $L^{(1)}(p_1,p_2,-p_3)$ \cite{Hernandez-Pinto:2015ysa,Rodrigo:prep}. 

\begin{figure}[ht]
\begin{center}
\includegraphics[width=5.5cm]{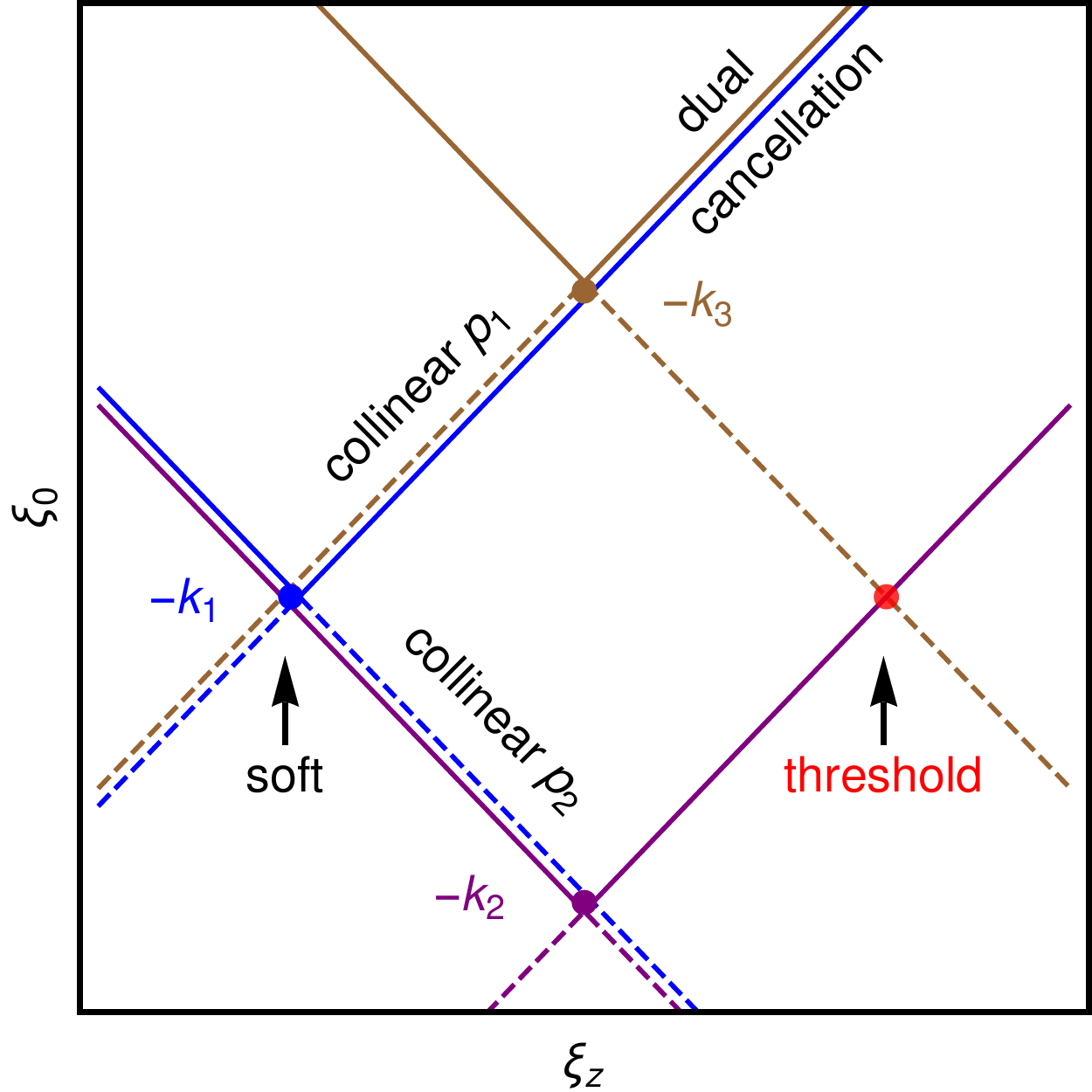}
\caption{Location of threshold and IR singularities in the $(\xi_0,\xi_z)$ space. Forward-forward (FF) singularities cancel among dual contributions whilst forward-backward (FB) intersections lead to IR and threshold singularities. In particular, the collinear singularities are spread along the FB intersections, and the soft one is associated with the intersection of the three light-cones.
\label{fig:IRsingularities}}
\end{center}
\end{figure}

Once we have written the dual contributions, we analyze the structure of their integrands to identify the origin of loop singularities. The integration domain is defined by the possitive-energy solutions of $G_F(q_i)^{-1}=0$ in the loop-momentum space, which are geometrically described by intersecting light-cones (LCs) when considering massless propagators. In Fig. \ref{fig:IRsingularities}, these domains are shown for each dual contribution; we distinguish between forward ($q_{i,0}>0$) and backward ($q_{i,0}<0$) LCs. The intersections among LCs produce different kinds of singularities, as discussed in Refs. \cite{Buchta:2014dfa,Buchta:2015xda}. For the massless triangle, the intersection among the three LCs originates a soft singularity, which is associated with double $\epsilon$-poles. On the other hand, forward-backward intersections lead to collinear singularities. For instance, the intersection among the forward LC of $I_1$ and the backward LC of $I_3$ originates single $\epsilon$-poles related with those introduced in the collinear limit $p_1 \parallel q_1$. Analogously, when the forward LC of $I_2$ intersects the backward LC for $I_1$, we obtain those singularities associated with the collinear region $q_1 \parallel p_2$. While forward-forward intersections cancel among dual contributions, the integrable threshold-singularity is associated with the intersection of the forward LC for $I_2$ and the backward LC for $I_3$. In that region, $q_2^2=0=q_3^2$ implies that two internal lines are on-shell and, moreover, they have positive energy; thus, the dual diagram factorizes into the product of two tree-level amplitudes.

To conclude this section, let's use the graphical information available in Fig. \ref{fig:IRsingularities} to identify the regions that contribute to the IR singular structure of the triangle. We define the following soft/collinear integrals;
\bea
I_1^{\rm (s)} &=& I_1(\xi_{1,0}\le w)  ~, 
\label{I1softORIGINAL}
\\ I_1^{\rm (c)} &=&  I_1(w \le \xi_{1,0}\le 1~; v_1\le1/2)  ~, 
\label{I1collinealORIGINAL}
\\ I_2^{\rm (c)} &=& I_2(\xi_{2,0}\le 1+w~; v_2\ge1/2)  ~,
\label{I2collinealORIGINAL}
\eea
where we introduced an arbitrary cut $w>0$ to deal with the threshold region. Performing an explicit computation, we find
\bea
\nn I^{\rm IR} &=& I_1^{\rm (s)} + I_1^{\rm (c)} + I_2^{\rm (c)} =  \frac{c_{\Gamma}}{s_{12}} \, \left(\frac{-s_{12}-\imath 0}{\mu^2}\right)^{-\ep} 
\\ &\times&\left[\frac{1}{\ep^2}+\ln{2}\ln{w}-\frac{\pi^2}{3}
-2\li{-\frac{1}{w}}+\imath \pi \ln{2}\right] + {\cal O}(\ep)~,
\label{IIRexpansion}
\eea
which implies that $L^{(1)}(p_1,p_2,-p_3) = I^{\rm IR} + {\cal O}(\ep^0)$, i.e. all the IR singular structure of the triangle is due to a compact region in the integration domain. This is a crucial property to achieve a local cancellation of IR singularities when adding the real corrections \cite{Hernandez-Pinto:2015ysa,Sborlini:2015uia,Rodrigo:prep}.


\section{Physical example: $\gamma * \to q \bar{q}$ at NLO}
\label{sec:examplephysical}
Let's combine real and virtual contributions through the application of LTD. The first implementation of this approach was presented in Ref. \cite{Hernandez-Pinto:2015ysa}, where we obtained the NLO corrections to a generic $1 \to 2$ decay in the context of a scalar theory. Here, we will briefly describe the procedure for computing NLO QCD corrections to $\gamma* \to q \bar{q}$. A more detailed presentation is available in Ref. \cite{Rodrigo:prep}.

In first place, we compute the renormalized one-loop correction to $\gamma*(p_3) \to q(p_1)+ \bar{q}(p_2)$, i.e.
\beq
\sigma_{\v}^{(1)} = \frac{1}{2s_{12}} \, \int d\Phi_{1\to 2} \, \left[ 2 {\rm Re} 
\la {\cal M}^{(0)}|\overline{\cal M}^{(1)} \ra 
- \left( \Delta \overline{Z}_2(p_1) + \Delta \overline{Z}_2(p_2)\right) |{\cal M}^{(0)}|^2
\right]~,
\label{Virtualqq}
\eeq
that also includes the self-energy corrections to all the external particles; this is a crucial point to achieve a fully local cancellation of IR singularities. Besides that, let's mention that the presence of a non-trivial structure in the numerator leads to UV divergent terms that must be absorbed into renormalization counter-terms. Using LTD and expressing the internal momenta $q_i^{\mu}$ in terms of $(\xi_{i,0},v_i)$, we obtain 
\beq
\sigma_{\v}^{(1)} = \sigma_{\v,1}^{(1)} + \sigma_{\v,2}^{(1)}+ \sigma_{\v,3}^{(1)} \, ,
\eeq
where $\sigma_{\v,i}^{(1)}$ are the \emph{renormalized dual cross-sections}.

On the other hand, the real correction corresponds to the process $\gamma*(p_3) \to q(p_1')+ \bar{q}(p_2')+ g(p_r')$, and is given by
\bea
\nn \sigma_{\r}^{(1)} &=& \sigma^{(0)}  \, 
\cgt \, \gs^2 \, C_F \, \left(\frac{s_{12}}{\mu^2}\right)^{-\ep} \,  
\int_0^{1} dy_{1r}'  \, \int_0^{1-y_{1r}'} dy_{2r}' \, (y_{1r}' \, y_{2r}' \,y_{12}' )^{-\ep}\, \, 
\\ &\times& \left[ 4\left( \frac{y_{12}'}{y_{1r}'\, y_{2r}'} - \ep \right)
+ 2(1-\ep) \left( \frac{y_{2r}'}{y_{1r}'} + \frac{y_{1r}'}{y_{2r}'} \right)  \right] \, ,
\label{Realqqg}
\eea
where $y_{ij}'\, s_{12}=2 p_i' \cdot p_j'$, $\sigma^{(0)}$ denotes the Born level cross-section and $\cgt$ is the PS volume factor in $d$-dimensions. Then, we separate the real-radiation PS into two regions which only contains one collinear configuration. Explicitly, we use the identity 
\beq
1=\theta(y_{2r}'-y_{1r}')+\theta(y_{1r}'-y_{2r}') \, ,
\label{eq:ThetaSplit}
\eeq
to split the three-body PS into two disjoint regions, $R_1=\{y_{1r}'<y_{2r}'\}$ and $R_2=\{y_{2r}'<y_{1r}'\}$. Implementing this separation at integrand level in \Eq{Realqqg}, we obtain
\beq
\sigma_{\r,i}^{(1)}=\sigma_{\r}^{(1)}(y_{ir}'<y_{jr}') \ \ \ \ , \ \ \ \ \ \ i,j =\{1,2\} \, ,
\label{SigmaRealSplit}
\eeq
which fulfill $\sigma_{\r}^{(1)}=\sigma_{\r,1}^{(1)}+\sigma_{\r,2}^{(1)}$. The following step is the introduction of a momentum mapping that allows to generate $1 \to 3$ on-shell kinematics using the variables $(\xi_{i,0},v_i)$. For instance, we define
\bea
&& p_r'^{\mu} = q_1^\mu~, \qquad p_1'^{\mu} = - q_{3}^\mu + \alpha_1 \, p_2^\mu 
= p_{1}^\mu - q_{1}^\mu + \alpha_1 \, p_2^\mu~, \nn \\ 
&& p_2'^{\mu} = (1-\alpha_1) \, p_2^{\mu}~,  \qquad 
\alpha_1 = \frac{q_3^2}{2q_3\cdot p_2}~, 
\label{MappingRegion1}
\eea
and express $y_{ij}'$ in terms of $(\xi_{1,0},v_1)$ according to
\beq
y_{1r}' = \frac{v_1 \, \xi_{1,0}}{1-(1 - v_1)\, \xi_{1,0}}~, \qquad 
y_{2r}' = \frac{(1-v_1)(1-\xi_{1,0}) \, \xi_{1,0}}{1 - (1-v_1)\,  \xi_{1,0}}~, \qquad
y_{12}' = 1-\xi_{1,0}~.
\label{EcuacionYprima1}
\eeq
This mapping is specially suited for $R_1$ because it properly describes the limit $y_{1r}'\to 0$, corresponding to the collinear configuration $q_1 \parallel p_1$. An analogous momentum transformation is obtained for $R_2$. Then, we use the corresponding mapping to rewrite $\sigma_{\r,i}^{(1)}$ in terms of $(\xi_{i,0},v_i)$. Since 
\bea
\theta(y_{2r}'-y_{1r}') \equiv {\cal R}_1(\xi_{1,0}, v_1) &=& \theta(1-2v_1) \, \theta\left(\frac{1-2v_1}{1-v_1}-\xi_{1,0}\right)\, ,
\label{Mappingconditions1} 
\\ \theta(y_{1r}'-y_{2r}')\equiv {\cal R}_2(\xi_{2,0}, v_2) &=& \theta\left(\frac{1-\sqrt{1-v_2}}{v_2}-\xi_{2,0}\right) \, ,
\label{Mappingconditions2} 
\eea
the regions $R_1$ and $R_2$ are mapped into subsets inside the integration domain of the dual cross-sections $\sigma_{\v,1}^{(1)}$ and $\sigma_{\v,2}^{(1)}$, respectively. So, we define $\tilde{\sigma}_{\v,i}^{(1)} = \sigma_{\v,i}^{(1)}(R_i)$ with $i=\{1,2\}$. Notice that it is not required to deal with $\sigma_{\v,3}^{(1)}$ because it does not contribute to the IR-singular structure of the virtual part. In other words, only the integrands of $\sigma_{\v,1}^{(1)}$ and $\sigma_{\v,2}^{(1)}$ are needed to locally regularize the divergent behavior of the real contribution in the regions $R_1$ and $R_2$, respectively.

Once the real and dual cross-sections are expressed using the same sets of variables, we proceed to combine them at integrand level. At this point, we realize that the isolation of the IR singularities of the loop integrals into a compact region of the parameter space becomes essential, because the real contribution has a finite integration domain. We conclude that
\beq
\sigma_{i}^{(1)} = \tilde{\sigma}_{\v,i}^{(1)}+{\sigma}_{\r,i}^{(1)} \ \ \ \ \ \  , \ \ \ \ \ \ \ \ \ \ i=\{1,2\} \ \ ,
\label{SigmaIfinitas}
\eeq
are \emph{finite integrals} in the limit $\epsilon \to 0$. Moreover, since the NLO corrections to the total cross-section are finite by virtue of KLN theorem, the sum of the dual contributions which are not used in the definition of $\sigma_{i}^{(1)}$, i.e.
\beq
\bar{\sigma}_{\v}^{(1)} = \left({\sigma}_{\v,1}^{(1)}-\tilde{\sigma}_{\v,1}^{(1)}\right)+\left({\sigma}_{\v,2}^{(1)}-\tilde{\sigma}_{\v,2}^{(1)}\right)+ {\sigma}_{\v,3}^{(1)} \, ,
\label{Remanente}
\eeq
is also finite when $\epsilon=0$. The most important point is that we can explicitly find four-dimensional representations for Eqs. (\ref{SigmaIfinitas}) and (\ref{Remanente}). For $\sigma_{i}^{(1)}$ it is straightforward to prove that the limit $\epsilon \to 0$ leads to a regular integrand. In the case of $\bar{\sigma}_{\v}^{(1)}$, there are some subtleties since it involves three different sets of variables. If we just implement a shift in the energy component and unify the angular variables, i.e. $(\xi_{i,0}+a_i,v_i)=(\xi_0,v)$, taking the limit $\epsilon \to 0$ would lead to missing finite parts. This is due to a mismatch in the collinear limits of the different dual integrands. To cure this behavior, we must use the same coordinate system to describe the three internal momenta $q_i^{\mu}$. Once we perform this change of variables, we can consider $\epsilon \to 0$ at integrand level and recover the same result obtained in DREG.

Finally, if we add the four-dimensional representations of Eqs. (\ref{SigmaIfinitas}) and (\ref{Remanente}), we obtain
\beq
\sigma^{(1)} = \sigma_{1}^{(1)} + \sigma_{2}^{(1)} + \bar{\sigma}_{\v}^{(1)} = 3 \, C_F \, \frac{\as}{4\pi} \, \sigma^{(0)} ~,
\label{totalNLO}
\eeq
which agrees with the well-known total cross-section at NLO in $\as$. We would like to emphasize that, following the procedure sketched here, the result shown in \Eq{totalNLO} is obtained through a \emph{purely four-dimensional implementation}. 


\section{Conclusions and outlook}
\label{sec:conclusions}
The loop-tree duality (LTD) theorem is a theoretical tool which allows to decompose loop-integrals in terms of dual contributions. These dual contributions are build using single cuts by invoking to a suitable modification in the $\imath 0$ prescription. In this article, we firstly applied LTD to analyze the IR singular structure of the massless triangle integral. This led us to the conclusion that soft and collinear poles are originated in a compact region of the integration domain.

After that, we center in the description of the NLO QCD corrections to the process $\gamma * \to q \bar{q}$. We showed a procedure to combine real and virtual contributions, and implement the computation considering the limit $\epsilon \to 0$ at \emph{integrand level}. This a distinctive aspect of our approach, because it is not trivial to find an integral representation which is compatible with commuting the limit $\ep \to 0$ and the integral. In other words, we know that the addition of real and virtual contributions should lead to finite results; i.e. using DREG, we \emph{integrate} and after that we take the limit $\epsilon \to 0$, because all $\epsilon$-poles cancel. Our claim is stronger, because we found an algorithm that allows to directly combine real and virtual terms \emph{before integration}, based in the dual decomposition of virtual amplitudes. As explained in Section \ref{sec:examplephysical}, the representation that we obtained is not only four-dimensional but also compatible with the commutativity of the limit $\epsilon \to 0$. And the essential component of this technique is the \emph{real-virtual mapping}, which allows to generate the real radiation kinematics from the Born level invariants, plus the spatial component of the loop momentum. In this way, singularities of the dual and real terms are mapped into the same points in the integration domain, leading to a fully local regularization.

In conclusion, this approach constitutes an alternative to the traditional subtraction method, with the appealing possibility of increasing the computational efficiency \cite{Rodrigo:prep,Buchta:2015wna}. Moreover, the four-dimensional representations that we found can be obtained using purely algebraic methods, which also might shed light into the mathematical structures behind scattering amplitudes and cross-sections.


\section*{Acknowledgments}
This work has been supported by the Research Executive Agency (REA) under the Grant Agreement No. PITN-GA-2010-264564 (LHCPhenoNet), by CONICET Argentina, by the Spanish Government and ERDF funds from the European Commission (Grants No. FPA2014-53631-C2-1-P, FPA2011-23778) and by Generalitat Valenciana under Grant No. PROMETEOII/2013/007.


\end{document}